\documentclass[referee]{raa}            

\usepackage{graphicx,times}             
\usepackage{natbib}
\usepackage{amssymb,amsmath}
\bibpunct{(}{)}{;}{a}{}{,}

\usepackage[a4paper=true,dvipdfm=true,pagebackref=true]{hyperref}
\hypersetup{colorlinks = true, linkcolor = green, anchorcolor = red, citecolor = blue, filecolor = red, pagecolor = red, urlcolor = red}

\begin{document}

\title{The distance to the giant elliptical galaxy M\,87 and the size of its stellar subsystem\,$^*$
}
\setcounter{page}{1}          
\author{N.A.~Tikhonov\inst{1}, O.A.~Galazutdinova\inst{1}, G.M.~Karataeva\inst{2}}

	   \institute{Special Astrophysical Observatory, Nizhnij Arkhyz, Karachai-Cherkessian Republic, 
		Russia 369167; {\it ntik@sao.ru}\\
		\and    
	Astronomical Institute, St. Petersburg State University, Petrodvorets, 198504 Russia {\it }\\        
		\vs\no
}

\abstract{Stellar photometry in nine fields around the giant elliptical
	galaxy M\,87 in the Virgo cluster is obtained from archival images
	of the Hubble Space Telescope. The resulting Hertzsprung--Russell
	diagrams show populated red-giant and AGB branches. The position
	of the tip of the red-giant branch (the TRGB discontinuity) is found
	to vary with galactocentric distance. This variation can be
	interpreted as the effect of metal-rich red giants on the
	procedure of the measurement of the  TRGB discontinuity or as a
	consequence of the existence of a weak gas-and-dust cloud around
	M\,87 extending out to~$10\arcmin$ along the galactocentric radius
	and causing $I$-band absorption of up  to~$0\fm2$ near the center
	of the galaxy. The TRGB stars located far from the M\,87 center
	yield an average distance modulus of  \mbox{$(m-M) =
		30.91\pm0.08$}, which corresponds to the distance of $D
	=15.4\pm0.6$~Mpc. It is shown that stars in the field located
	between  M\,86 and M\,87 galaxies at angular separations of
	$37\arcmin$ and $40\arcmin$ are not intergalactic stars, but
	belong to the M\,87 galaxy, i.e., that the stellar halo of this
	galaxy can be clearly seen at a galactocentric distance of
	$190$~kpc. The distances are measured to four dwarf galaxies
	P4anon, NGC\,4486A, VCCA039, and dSph-D07, whose images can be
	seen in the fields studied. The first three galaxies are  M\,87
	satellites, whereas dSph-D07 is located at a greater distance and
	is a member of the  M\,86 group.
}

\authorrunning{N.A.~Tikhonov, O.A.~Galazutdinova, G.M.~Karataeva}            
\titlerunning{The distance to the giant elliptical galaxy M\,87 }  

\maketitle

\section{Introduction}           
\label{sect:intro}

As a result of the gravitational interaction between neighboring
galaxies in the densely populated Virgo cluster stellar
peripheries should be "stripped" from galaxies and intergalactic
stars should appear~\citep{Rudick_2006}. Interaction
processes should also alter the morphology of the stellar
subsystems of galaxies, especially their extended halos. However,
our analysis of galaxies in several fields of the Virgo cluster
whose images were taken with the Hubble Space Telescope revealed
that the stellar subsystem of each dwarf galaxy extends out to
large galactocentric distances~\citep{Tikhonov:2017} and
occupies the entire area on the corresponding images. Finding
likely intergalactic stars among numerous stars of the periphery
of the galaxies, proving their intergalactic status, and
determining their parameters is an extremely challenging task.

Deep Hubble Space Telescope images have been obtained for very few
fields in the Virgo cluster. One of these fields, which \citet{Williams_etal:2007a} analyzed to search for
and study intergalactic stars, is located between the two giant
elliptical galaxies M\,86 and M\,87 at the angular separations of
37\arcmin\ and~40\arcmin\,  respectively, which correspond to
about~170 and 190~kpc for the adopted Virgo cluster distance of
16.5~Mpc. The location of this field was chosen based on an
analysis of deep Virgo cluster
images~\citep{Mihos_etal:2005} taken with the
0.6\mbox{-}m Schmidt telescope and reaching the 29$^{\rm
m}$~arcsec$^{-2}$ $B$-band isophote. These and later images
obtained by the same authors~\citep{Mihos_etal:2017}
show that the considered  field is located beyond the extended
halos of the two neighboring galaxies M\,86 and M\,87. However,
\citet{Janowiecki_etal:2010} analyzed
the same images~\citep{Mihos_etal:2005} and found that
the halo of  M\,87 extends out to~40\arcmin, i.e., that the field
in question should be located within the M\,87 halo.

While studying the images obtained with the Hubble Space Telescope 
\citet{Williams_etal:2007a} took into
account the conclusions of ~\citet{Mihos_etal:2005} and assumed that the field
considered was located far enough from the giant galaxies for
their contribution to the star sample to be negligible, and
therefore attributed the inferred stellar parameters to
intergalactic stars of the Virgo cluster.

The discovery of intergalactic stars in the Virgo cluster was
reported earlier by~\citet{Caldwell:2006}, who
studied two fields in  the images obtained with the ACS/WFC camera
of the Hubble Space Telescope. We analyzed the same observational
data and concluded that intergalactic stars are totally lacking or
extremely scarce~\citep{Tikhonov:2017}. We therefore
considered it necessary to investigate the field
from~\citet{Williams_etal:2007a} in order to
independently validate the presence of intergalactic stars in it
and study their parameters.

\begin{figure*}[bpt!!!]
	
\centering
\includegraphics[width=13cm,clip]{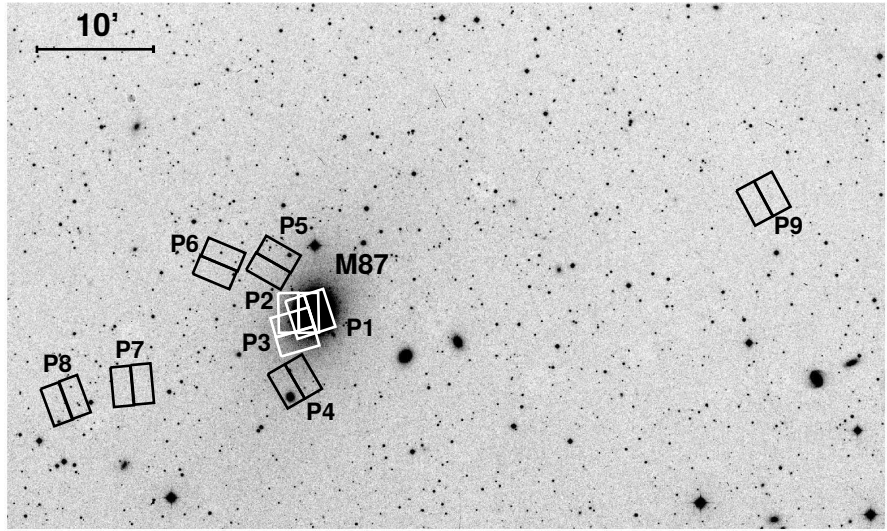} 
\caption{Image of the part of the Virgo cluster with the M\,87
galaxy in the DSS photo. The squares indicate fields~\mbox{P1--P9}
as observed with the ACS/WFC camera of the Hubble Space Telescope.
North is at the top.}\label{fig1}
\end{figure*}

\citet{Tikhonov:2017} showed that the
distribution of stars of neighboring galaxies has to be studied to
determine whether stars in a certain field are members of a
certain galaxy. The surface photometry method used
in~\citet{Mihos_etal:2005} to determine the structure of
haloes around galaxies cannot be applied to isophotes fainter than
29--30$^{\rm m}$ and does not make it possible to determine the
type of stars that form the particular halo. We therefore use the
method of star counts and complement the study of the field of \citet{Williams_etal:2007a}, which denote
here as P9,  by investigating eight more fields around the  M\,87
galaxy that are resolvable into stars~(Fig.~\ref{fig1}) in order to study
the variation of the parameters of stars with galactocentric
distance. The results obtained led us to a conclusion about the
nature of stars in field~P9, which we report in this paper.

According to  the NASA Extragalactic Database\footnote{\url
{https://ned.ipac.caltech.edu/}} (NED), the elliptical galaxy
M\,86 is  fainter than M\,87 and has a smaller-sized stellar
subsystem. However, field~P9~(Fig.~\ref{fig1}) is located along the major
axis of M\,86, where the size of its stellar halo is the greatest.
Therefore stars in the considered  field can be both intergalactic
objects or belong to the  periphery of the  M\,86 and M\,87
galaxies. The ratio number of stars belonging  to different
galaxies in this field can be measured by performing stellar
photometry of this field and the neighboring galaxies.

The principle indicator of the membership of a star in the considered field
 in a particular galaxy is the equal distance to both
the galaxy and to the stars of the considered field (provided that
the stars are not broadly distributed along the line of sight).
The TRGB (Tip of Red Giants Branch)
method~\citet{Lee_etal:1993}, which  allows the distance
to a group of red giants to be determined sufficiently accurately
and reliably is probably the only method for determining the
distance in the considered case.

\section{Choice of fields and stellar photometry}
\begin{figure*}[bpt!!!]
	\centering
\includegraphics[width=15cm,clip]{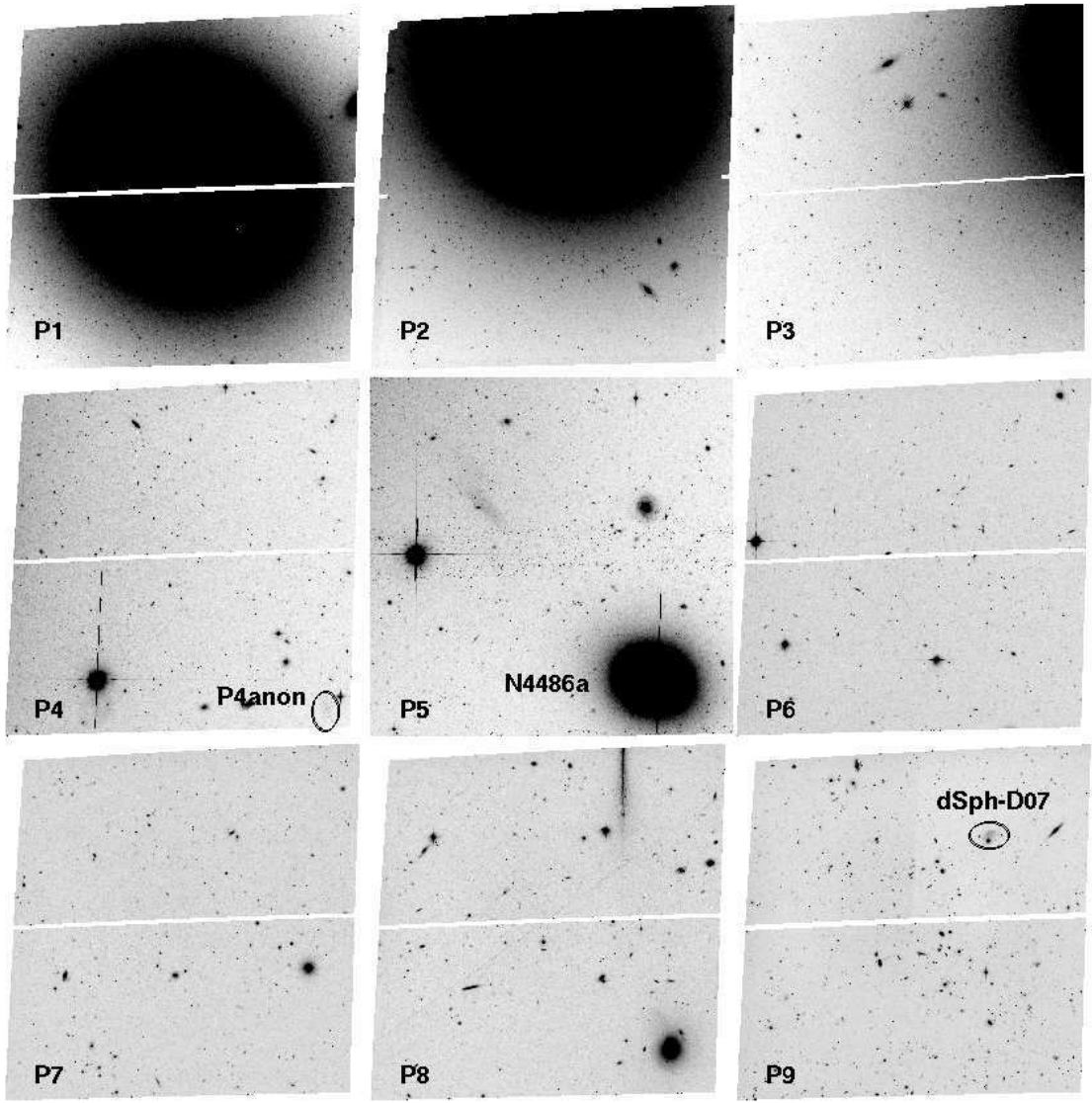}
\caption{Fields P1--P9 on Hubble Space Telescope images. The
ellipses in fields P4 and P9 indicate dwarf galaxies. All images
have the size of $3\farcm5\times3\farcm5$.}\label{fig2}
\end{figure*}

To study the space structure of stellar subsystems in the M\,86
and M\,87 galaxies of the Virgo cluster and determine accurate
distances and study intergalactic stars, we used archival
images taken by the Hubble Space Telescope (HST) within the
following applications: ID\,10131, 10543, 12532, 12989, and 13731.
The main criterion for the choice was that the $F814W$~($I$)-band
ACS/WFC exposure should be 2000 seconds or longer. We did not
require the availability of images of the same field taken with
another filter, although we consider such images to be useful. We found 
 deep images of a total of nine fields, which we hereafter
refer to as \mbox{P1--P9}, located at different distances
from~M\,87. We also found four fields for M\,86, however,
photometry showed that this galaxy is 2.0--2.5~Mpc farther way
than stars of field P9, and we therefore do not consider it in
this study. We show the location of fields P1--P9 with respect to
the M\,87 galaxy in the DSS (Digital Sky Survey) image in Fig.~\ref{fig1},
and HST archival images in Fig.~\ref{fig2}.

\begin{table*} 
	\begin{center}
		\caption{Exposures of P1---P9 field images (in seconds) } \label{tab1}
		\medskip
		\begin{tabular}{c|c|c|c|c|c|c|c|c|c}
			\hline\noalign{\smallskip}
			Filter & P1 & P2 & P3 & P4 & P5 & P6 & P7 & P8 & P9 \\
			\hline\noalign{\smallskip}
			$F814W$~($I$)& 73800 & 4270 & 2589 & 2282 & 12860 & 2282 & 2282 & 2282 & 26880\\
			$F606W$~($V$)& 24500 & --   & --   & --   & 12050 & --   & --   & --   & 63440\\
			$F475W$~($B$)& --    & --   & 2729 & 2351 & --    & 2351 & 2351 & 2351 & --   \\
			\hline
		\end{tabular}
	\end{center}
\end{table*}

We should point out the heterogeneity of the data used. For fields
P1, P5, and P9 we have deep $F814W$~($I$) and $F606W$~($V$)-band
images, and this is optimal for distance measurement. Less deep
$F814W$~($I$) and $F475W$~($B$)-band images are available for the
other fields (see Table~\ref{tab1}, which summarizes the data about image
exposures and filters). The distances determined using the TRGB
method rely on red giants, which are near the photometric limit in the
images taken with the blue filter $F475W$~($B$), but are quite
conspicuous in the images taken with  the $F814W$~($I$) filter.
Combined photometry of  $F814W$- and $F475W$-band images, as it is
done in DOLPHOT package~\citet{Dolphin:2016}, sets the
common photometric limit based on the less deep  $F475W$-band
image.
DAOPHOT~II~\citep{Stetson:1987,Stetson:1994}
package of MIDAS provides sufficient amount of stellar photometry
in the deepest  $F814W$-band filter, allowing the  TRGB
discontinuity to be determined from deeper images, although the
lack of star color information somewhat complicates the
identification of red giants.

We performed stellar photometry in a standard way using DAOPHOT~II
software, as described in our earlier
paper~\citet{Tikhonov_etal:2009}. For aperture
photometry we used smaller than recommended standard aperture
sizes (Table~\ref{tab2}) to ensure more crowded arrangement and more
precise description of the photometric profile at the centers of
the stars.

\begin{table*}
\begin{center}
	\caption[]{Aperture sizes used for DAOPHOT~II photometry of ACS/WFC
HST images}\label{tab2}
\medskip
\begin{tabular}{c|c|c|c|c|c|c|c|c|c|c|c|c|c|c} \hline\noalign{\smallskip}
Aperture& A1 & A2 & A3 & A4 & A5 & A6 & A7 & A8 & A9 & AA & AB &
AC & IS & OS
\\
\hline\noalign{\smallskip}
Radius, pixels&1.5 & 1.7 & 2.0 & 2.4 & 2.9 & 3.5 & 4.2 & 5.0 & 5.8 & 7.9 & 8.5 & 10.0 & 12 & 20\\
\hline\noalign{\smallskip}
\end{tabular}
 \end{center}
\end{table*}

We selected the results of our stellar photometry by parameters
``$CHI$''$< 1.5$ and ``$|SHARP|$''$< 0.3$, which determine the shape
of the photometric profile of each measured
star~\citep{Stetson:1987}, making it possible to
eliminate from photometric tables all diffuse objects---star
clusters, distant or compact galaxies---because profiles of these
objects differ from those of  isolated stars selected as
standards.

We used  DOLPHOT 2.0 package\footnote{\url
{http://americano.dolphinsim.com/dolphot/}} in accordance with the
recommendations given in the manual. The procedure of photometry
consisted of preliminary masking of ``bad'' pixels, cosmic-ray hit
removal, and subsequent PSF photometry of extracted stars in two
filters. Selection of stars from the resulting preliminary list by
image-profile parameters ``$CHI$''  and  ``$|SHARP|$'' was performed
in the same way as in DAOPHOT~II. To perform photometry of stars
in the central region of the M\,87 galaxy (field P1), we aligned
and averaged more than 200 images. However, despite very long
exposure, the photometric limit of the images in the central part
of  this field is much shallower because of the strong crowding of
stars and the brightness of the galaxy.

The use of the sole filter ($F814W$) for determining the TRGB
discontinuity and computing the distance to a galaxy has its
certain unique features. Given that it is impossible to
obtain a Hertzsprung\mbox{--}Russell diagram (the CM diagram) to
identify red giants by color, the resulting luminosity function
refers to all field objects. If the field contains many metal-poor
red giants then the TRGB jump is always quite conspicuous even in
the case where an image taken in one filter is used. However, if
the star sample consists of a mix of metal-rich and metal-poor
stars, which is the case in elliptical galaxies, the TRGB
discontinuity shows up only as the variation of the gradient of
the luminosity function. In many cases such a variation of the
gradient is more conspicuous if logarithmic scale is used for the
number of stars.

We used the following equations to transform instrumental
magnitudes into the Kron--Cousins $V$- and $I$-band magnitudes in
DAOPHOT~II:
    $$(V-I) = 1.3213(v-i) + 1.133, \eqno(1)$$
    $$I = i + 0.0592(V-I) + 25.972, \eqno(2)$$

where $(v-i)$ and $i$ are the instrumental magnitudes and $(V-I)$
and $I$ are the magnitudes in the corresponding passbands of the
Kron--Cousins system. We derived the above equations based on
photometry of the same stars carried out out with different
telescopes and detectors: 6-m telescope of the Special
Astrophysical Observatory of the Russian Academy of Sciences, 1-m
Zeiss-1000 telescope, and the Hubble Space Telescope with WFPC2
camera~\citep{Tikhonov_Gala:2009}. The transformation
equations are accurate to within  $0\fm02$ and $0\fm03$ for
$I$-band magnitude and  $(V-I)$ color index, respectively.

The second equation  shows a weak dependence of the resulting
$I$-band magnitude on  $(V-I)$ color index. Hence the use of the
average color index $(V-I) = 1.6$, which is close to real colors
of red giants in elliptical galaxies, should result in extra error
of only $0\fm01$ for stars that are  $0\fm2$ bluer or redder. It
is clear that such transformation cannot be used for simultaneous
reduction of instrumental magnitudes to $I$-band magnitudes for
blue and red stars, however, no blue stars are present in the
fields where photometry was performed.

Photometry in fields P2\mbox{--}P9 was performed with  DAOPHOT~II,
and photometry in fields P1, and P4\mbox{--}P9 was additionally
performed using  DOLPHOT package. The accuracy of stellar
photometry performed in the~$F814W$ filter is given in Table~\ref{tab3}.

\begin{table}
	\begin{center}
\caption[]{Accuracy of $F814W$-band stellar photometry in
fields~\mbox{P1--P9} around  M\,87}\label{tab3}
\medskip
\begin{tabular}{c|c|c|c|c|c|c|c|c|c} \hline\noalign{\smallskip}
$m$*& P1 & P2 & P3 & P4 & P5 & P6 & P7 & P8 & P9 \\
\hline\noalign{\smallskip}
25 & 0.02 & 0.05 & 0.03 & 0.02 & 0.04 & 0.03 & 0.03 & 0.05 & 0.02\\
26 & 0.04 & 0.08 & 0.05 & 0.04 & 0.07 & 0.06 & 0.05 & 0.07 & 0.02\\
27 & 0.12 & 0.15 & 0.10 & 0.13 & 0.14 & 0.14 & 0.12 & 0.14 & 0.03\\
28 & 0.30 & 0.35 & 0.25 & 0.32 & 0.30 & 0.32 & 0.28 & 0.33 & 0.07\\
\hline\noalign{\smallskip}
\end{tabular}
\begin{tabular}{lp{80mm}}
* $m$ --- apparent magnitude in the $F814W$ filter
\\[-5pt]
\end{tabular}
 \end{center}
\end{table}

We determined Galactic foreground extinction towards fields P1--P9
in accordance with ~\citet{Schlafly_Finkbeiner:2011}. Extinction
is small towards all fields and varies from $A_I = 0\fm040$ for
field P8 to \mbox{$A_I = 0\fm032$} for field P4.

\section{Stellar subsystems M\,87}

\begin{figure}[]
	\centering
	\includegraphics[width=0.45\columnwidth]{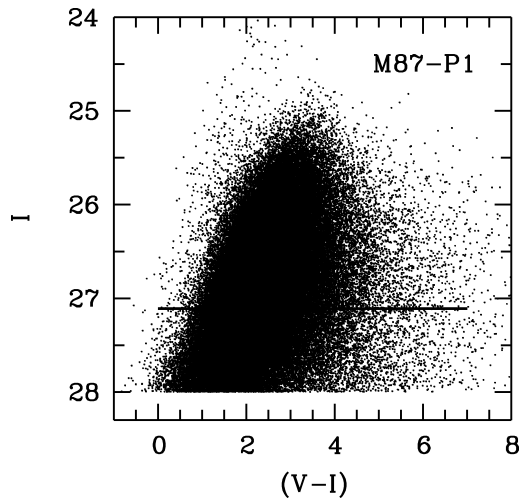}
\caption{Color-magnitude diagram for the central field (P1) of the
M\,87 galaxy. The horizontal line indicates the upper boundary of
the red-giant branch---the TRGB discontinuity. The domain of very
bright AGB stars is located above the TRGB discontinuity.}\label{fig3}
\end{figure}

\begin{figure}[]
	\centering
\includegraphics[width=0.45\columnwidth]{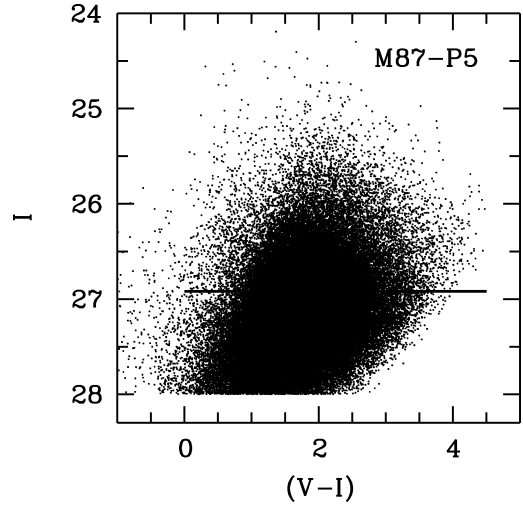}
\caption{Color-magnitude diagram for stars in field P5. The
horizontal line indicates the location of the TRGB discontinuity.
Very bright AGB stars are practically absent, but there are quite
a few intermediately bright AGB stars.}\label{fig4}
\end{figure}

\begin{figure}[]
	\centering
	\includegraphics[width=0.45\columnwidth]{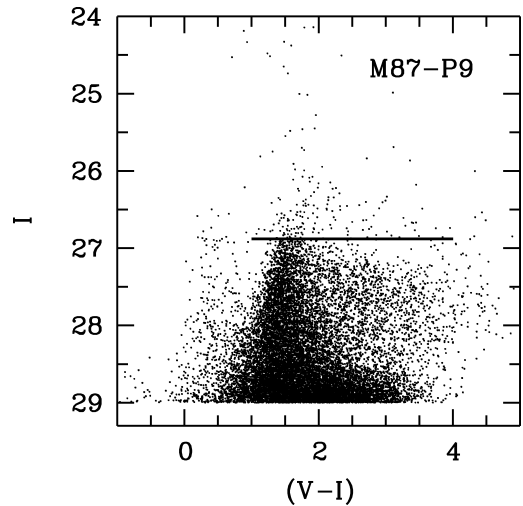}
\caption{Color-magnitude diagram for stars in field P9. Almost
vertical branch of metal-poor red giants can be seen at
\mbox{$(V-I) = 1\fm5$}. The horizontal line indicates the TRGB
discontinuity on this branch. Metal-rich red giants are located in
the domain with magnitudes from \mbox{$I = 27\fm2$} to \mbox{$I =
28\fm5$} and color indices \mbox{$(V-I)$} from~$2\fm0$ to~$4\fm5$.
A small number of low-luminosity AGB stars are located above the
TRGB discontinuity.}\label{fig5}
\end{figure}

Figures~\ref{fig3},~\ref{fig4}, and~\ref{fig5} show the color-magnitude diagrams of fields
P1, P5, and P9. The diagram for field P1 Fig.~\ref{fig3}) contains many
very bright AGB stars located above the TRGB discontinuity, which
for this field is equal to  $I$ = 27.10. Given that such a diagram
is typical for any bright elliptical galaxy it is absolutely
unclear how ~\citet{Bird_etal:2010}, based on
their photometry of the same images, left only red giants on their
color-magnitude diagram removing all brighter AGB stars. Besides
AGB stars, the color-magnitude diagram in Fig.~\ref{fig3} contains many
metal-rich red giants whose color indices extend  to  $(V-I) =
6\fm0$. The color-magnitude diagram of field P5, which is farther
away from the center of the galaxy than field P1, shows numerous
AGB stars but no very luminous stars like those in field P1 among
them. Field P9 is the photometrically deepest among the fields
considered. Its color-magnitude diagram features the branch of
metal-poor red giants and also a small domain of metal-richer red
giants. There are very few AGB stars in this field. The weak
branch of blue objects at $(V-I) = 0\fm4$ is made up of foreground
objects and is not associated with M\,87. Stars in field P9 have
the most accurate photometry and hence yield the most precise
measurement of the position of the TRGB discontinuity and the
distance to M\,87.

\begin{figure}[]
		\centering	
\includegraphics[angle=0,width=6cm,clip]{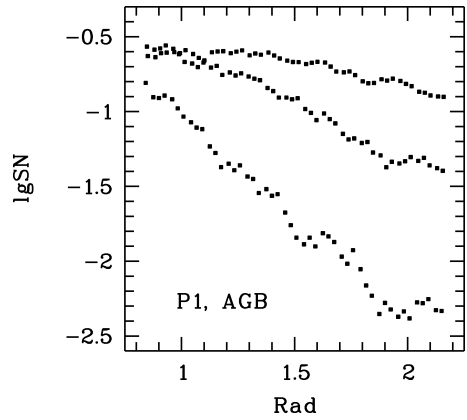}
\caption{Distribution of the
number density of AGB stars of different luminosity along the
galactocentric radius in the central field P1. The higher the
luminosity of stars, the higher is the gradient of the density
decrease. Then horizontal axis gives the angular distance from the
center of M\,87 in arcmin.}\label{fig6}
\end{figure}

Figure~\ref{fig6}, which shows the variation of the number density  of AGB
stars of different luminosity along the galactocentric radius in
field P1. It is evident from the figure that the higher the
luminosity of stars the steeper is the gradient of the decrease of
the number density. The behavior stars in accordance with the
diagram in Fig.~\ref{fig6} means that no very bright AGB stars can be seen
relatively close to the center of M\,87. At large galactocentric
distances even intermediately-bright AGB stars become scarce and
the color-magnitude diagrams in Figs.~\ref{fig3},~\ref{fig4},~\ref{fig5} fully corroborate this
fact.

\begin{figure}[bpt!!!]
	\centering
\includegraphics[angle=0,
width=8.0cm,clip]{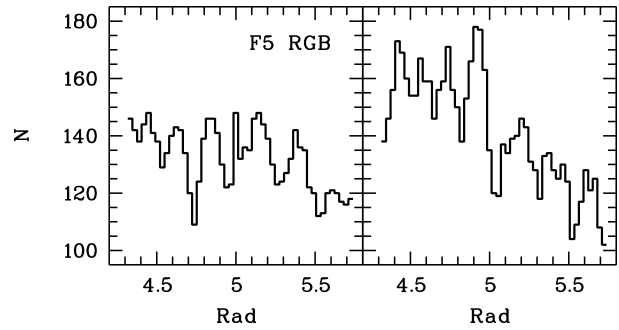} \caption{Distribution of
the number density of red giants of different metallicities along
galactocentric radius in field P5. Metal-poor red giants show
lower radial density decline gradient than higher-metallicity
giants, and this results in the variation of morphology of the
star sample along galactocentric radius. The horizontal axis gives
the distance from the center of M\,87 in acrmin.}\label{fig7}
\end{figure}

Figure~\ref{fig7} shows the diagram of the variation of the number density
of metal-poor and metal-rich red giants along galactocentric
radius. As is evident from the figure, metal-poor giants show a
less steep distribution along galactocentric radius. This
dependence results in the variation of the proportion of
metal-poor and metal-rich red giants along galactocentric radius
in M\,87. None of the dependences found shows any abrupt change of
behavior with stellar metallicity. This means that the space
distribution laws for intermediate-metallicity stars should be
intermediate between those of metal-poor and metal-rich stars.

The differences between the gradients of the decline of the number
density of stars of different types result in the variation of
color index along the radius. \citet{Montes_etal:2014} used such photometric
variations to compute the metallicity and age of stars along
galactocentric radius. This method allows us to trace the color
variations in the central region of M\,87, but it remains
``blind'' for interpreting the population type in this region
because instruments detect only the sum of the intensities of
stars of all types in the region considered without making it
possible to separate, e.g., the contributions of high- and
low-luminosity AGB stars because they have almost identical color
indices.

\begin{figure}[bpt!!!]
	\centering
\includegraphics[angle=0,
width=8.0cm,clip]{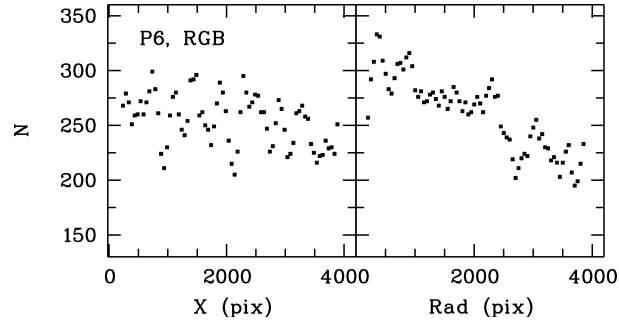} \caption{Distribution of
the number density  of red giants in field P6 perpendicularly to
the radial  direction of M\,87 and along this direction. The small
gradient of the distribution in panel (a) can be explained by the
not  very accurate positioning of field P6 along the radial
direction. }\label{fig8}
\end{figure}

\begin{figure}[bpt!!!]
\centering
\includegraphics[angle=0,
width=8.0cm,clip]{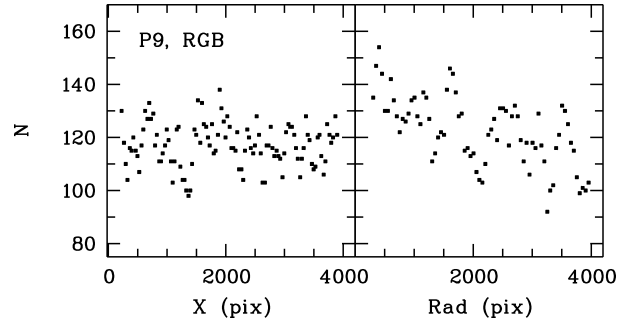}
 \caption{Same as in Fig.~8,
but for field P9. The observed uniform distribution of the number
of stars in the direction perpendicular to the radius of M\,87
(the X-axis) and the increase of the number of stars along this
direction indicate that stars in field P9 belong to the periphery
of the M\,87 galaxy.}\label{fig9}
\end{figure}

The  M\,87 membership of stars in fields P1--P5 is beyond question
given the presence of the galaxy in the images, but M\,87 cannot
be seen in fields P6\mbox{--}P9. Can stars in these fields belong
to other galaxies? Partially can, because, e.g., in fields P7 and
P9 there are two dwarf galaxies with our distance estimates.
However, the bulk of the stars in fields P6--P9 belong to M\,87.
We constructed the distribution of the number density of red
giants toward the direction perpendicular to the radius of M\,87,
and the corresponding distribution along the radius for field P6
and for the outermost field P9 (Figs.~\ref{fig8} and \ref{fig9}). The increase of
the number of stars toward M\,87 and the uniform distribution  of
these stars in the perpendicular direction are immediately
apparent. These diagrams combined with the fact that the distances
to different fields around  M\,87 are the same, as we will show
below, prove conclusively the membership of AGB stars and red
giants in these fields in extended stellar subsystems M\,87.
Projections of dwarf galaxies introduce minor distortions into
this global structure, but this does not alter the general
conclusion that the stellar periphery in M\,87 extends out to a
radius of $40\arcmin$ and appears to extend even farther judging by
the number of stars in the outermost field.

Massive galaxies are known to have extended stellar
halos~(\citealt{Tikhonov_etal:2005,Rudick_etal:2010,Rejkuba_etal:2014,Ibata_etal:2014}),
which are observed around the galaxies either using the methods of
surface photometry or by measuring the number density of stars,
provided that the galaxy can be resolved into stars. Measurements
of some giant galaxies revealed that they possess extensive
stellar halos. \citet{Rejkuba_etal:2014}
traced the extent of the stellar periphery out to 140~kpc in the
NGC~5128 galaxy in Centaurus cluster, which is less luminous than
the galaxies M\,86 and M\,87 in Virgo cluster.
 \citet{Mihos_etal:2013} performed surface
photometry to find the stellar halo in the M\,49 galaxy extending
out to a galactocentric distance of 100~kpc.
\citet{Kormendy_etal:2009} estimated the
size of the stellar periphery of M\,87 and found its stellar
component to extend out to 200~kpc. \citet{Oldham_Auger:2016} used the results of
\citet{Kormendy_etal:2009} in their
proposed model of the structure of M\,87 and estimated the size of its
stellar component to be 1~Mpc. Naturally, no real measurements can
confirm or disprove this estimate because the presence of bright
neighboring galaxies makes it impossible to measure the extremely
faint halo at such a galactocentric distance even if is actually
present there.

Note that the surface brightness method can be used in clusters of
galaxies only down to a certain isophote. Because of the high
concentration of galaxies, effects arising when different galaxies
located far apart are projected onto the same field, and possible
glow of intergalactic objects:
stars\citep{Ferguson_etal:1998,Durrell_etal:2002,Williams_etal:2007a},
planetary
nebulas\citep{Arnaboldi_etal:2002,Mihos_etal:2009,Longobardi_etal:2015},
and globular
clusters~\citep{Williams_2007b,Ko_etal:2017},
moving to very deep brightness levels would result in the
isophotes of all objects merging into a single large cloud. In
this case it will be impossible to  determine the boundaries of
faint isophotes of individual objects. The method of star counts
for measuring the sizes of galaxy halos is barely affected by
background and foreground objects, but it requires deep images,
which cannot always be acquired.

\section{Measurment of distance}
The distances to the  M\,86 and M\,87 galaxies have been measured
repeatedly using different methods.
The NED database 
contains more than one hundred distance estimates. The mean and
median  distance estimates are equal to 16.08 and 16.80~Mpc,
respectively, for M\,86  and 16.56 and 16.40~Mpc, respectively,
for M\,87 (NED). Some methods, e.g., the one based on planetary
nebulas, yield smaller-than-average estimates (14.6~Mpc for
M\,87), and the method based on the brightness of novas yields
greater-than-average estimates (18.8~Mpc for M\,87). This appears
to be due to the calibration of the respective zero points and the
extensive program aimed at the study of novas in
M\,87 (\citealt{Shara_etal:2016,Shara_etal:2017}) will change the situation.

The NED database contains the results of measurements made using all
methods with different accuracy, and
therefore to determine the true distances we should rely on
the TRGB method, which is currently considered to be the most
precise among the methods applied for elliptical galaxies. This
method yielded the following distance estimates for M\,87:
15.1~Mpc~\citep{Lee_Jang:2016a},
15.2~Mpc~\citep{Ferrarese_etal:2000},
16.0~Mpc~\citep{Lee_Jang:2016b},
16.7~Mpc~\citep{Bird_etal:2010},
19.4~Mpc~\citep{Ferrarese_etal:2000}. If we reject the
evident outlier---19.4~Mpc---then the average distance estimate
becomes  15.75~Mpc. Note that the TRGB method-based  distance
measurements in the above papers in each case relied on the data
for a single field. Such measurements are quite sufficient for
most of the galaxies, but, as our measurement showed, for M\,87 several fields
 located at different galactocentric distances have to be used.

No TRGB-based distance measurements for M\,86 are available in
NED, but our earlier estimates~\citep{Tikhonov:2017} as
well as new distance estimates based on four fields indicate that
M\,86 is located about 2~Mpc farther than  M\,87. If field P9
located between M\,86 and M\,87 contains stars of both galaxies
then  we have to observe two separate TRGB discontinuities on the
luminosity function of red giants.

The color-magnitude diagrams of M\,87 fields (Figs.~\ref{fig3},~\ref{fig4},~\ref{fig5})
contain red giants, which at galactocentric separations smaller
than  10\arcmin\ are masked by brighter AGB stars. The presence of
red giants makes it possible to use the TRGB method in accordance
with ~\citet{Lee_etal:1993} for distance
measurement. To determine the position of the upper tip of the red
giant branch we applied the Sobel
filter~\citep{Madore_Freedman:1995} to the luminosity
function of red giants and AGB stars. The maxima of  the Sobel
filter indicate the locations of abrupt change of the luminosity
function gradient, which is  observed when passing from AGB stars
to the beginning of the red-giant branch. The TRGB discontinuity
is hardly visible in the case of giant elliptical galaxies because
the sample studied includes stars of different metallicity. In
this case, the variation of gradient of the luminosity function is
observed instead of the TRGB discontinuity. The position of the
TRGB discontinuity in the $I$-band filter depends slightly on the
metallicity of red giants, and therefore it is necessary to know
the metallicity of these stars. In the method of~\citet{Lee_etal:1993} metallicity is measured via
two quantities: $(V-I)_{\rm TRGB}$, the color index of the tip of
the red giant branch, and $(V-I)_{-3.5}$, the color index of the
red-giant branch at the  $M_I = -3.5$ level. Once the location of
the TRGB discontinuity on the luminosity function  and the color
indices of the red-giant branch on the color-magnitude diagram are
determined, the distance to the galaxy studied can be computed as
described in~\citet{Lee_etal:1993}.

\begin{figure*}[bpt!!!]
\centering
\includegraphics[width=0.8\textwidth]{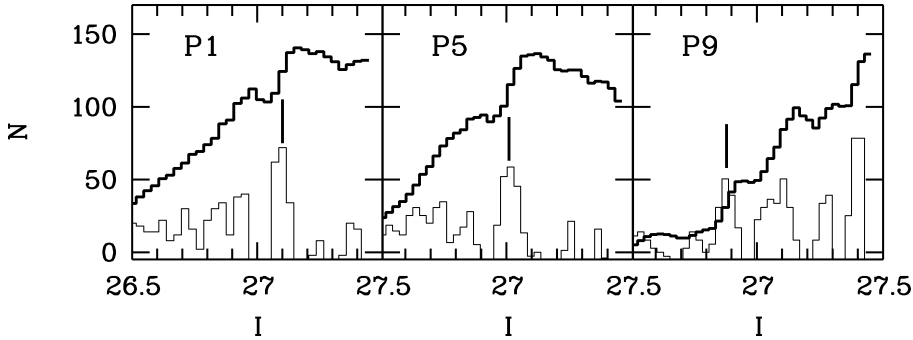}
\caption{Luminosity functions of stars near the boundary of the
red-giant branch for fields P1, P5, and P9. Because of the
presence of stars of different metallicities in the sample the
TRGB discontinuity is observed on the luminosity function shows up
as the variation of the gradient. The thin line shows the Sobel
function, which indicates maximum gradients of the luminosity
function. The vertical bars indicate the positions of TRGB
discontinuities, which for the given sample correspond to the
boundary of red giants.}\label{fig10}
\end{figure*}

\begin{figure*}[bpt!!!]
	\centering
\includegraphics[width=0.8\textwidth]{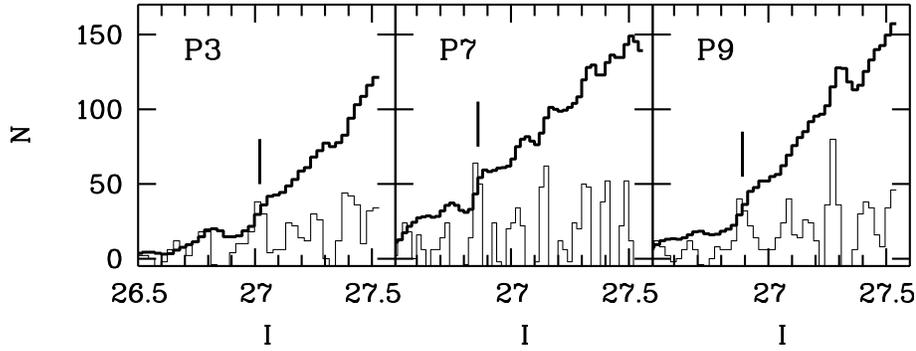}
\caption{Luminosity functions for fields P3, P7, and P9. Unlike
Fig.~\ref{fig10}, where the sample of stars was subjected to~\mbox{($V-I$)}
color cut, the luminosity functions presented here are based on
photometry in the sole~$I$-band filter. A comparison of the
diagram for  field P9 with the corresponding diagram in Fig.~\ref{fig10}
shows their similarity, i.e., confirms the possibility  of
measuring the TRGB discontinuity based on  stellar photometry in
the sole~$I$-band filter .}\label{fig11}
\end{figure*}

Figure~\ref{fig10} presents the luminosity functions for stars in fields
P1, P5, and P9. Despite long exposures in field P1 the TRGB
discontinuity can be seen only on the luminosity function of stars
located farther than  \mbox{Rad\,$>90\arcsec$} from the center of
the galaxy. The central part of M\,87 is so bright and overcrowded
with stars that red giants are lost amongst brighter and more
numerous AGB stars. The change of the gradient of the luminosity
function for stars of this field (the TRGB discontinuity) is
observed at \mbox{$I = 27.10$}.
The luminosity function for stars of field P5, located farther away from the center of the galaxy, 
changes its gradient at  I=27.01, which is the TRGB discontinuity. For an even more distant field P9, a change in the gradient of the luminosity function can be seen at I = 26.89.

Figure~\ref{fig15} shows the luminosity functions for fields P3, P7, and
P9. The luminosity function for field P9 is based on the results of
single-band ($I$) stellar photometry. A comparison with Fig.~\ref{fig10},
where a similar luminosity function is based on photometry on two
filters and selection of red giants based on the color-magnitude
diagram, shows no difference in the resulting measured TRGB
discontinuity. Hence images taken with the sole $I$-band filter
are quite sufficient for determining the TRGB discontinuity,
provided, naturally, that the possible  presence of stars of other
types than red giants is taken into account.
\begin{figure}[bpt!!!]
	\centering
\includegraphics[width=0.8\columnwidth]{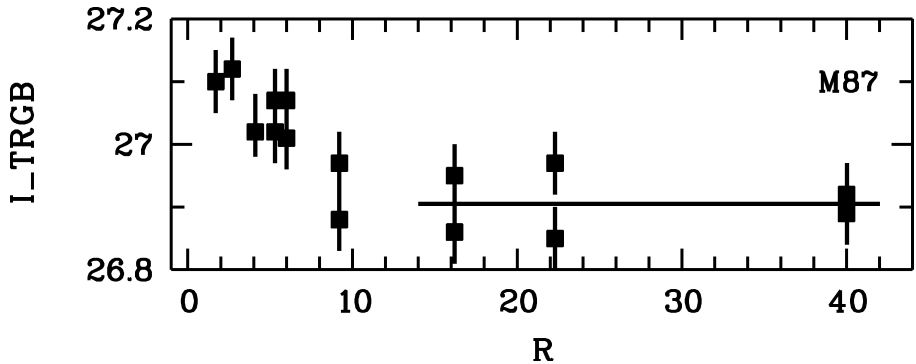} 
\caption{Variation of the
position of the TRGB discontinuity on the stellar luminosity
function with the distance of the considered field  from the center
of the galaxy. We appear to see the effect of metal-rich red
giants whose number decreases with the distance from the center of
the galaxy. No  systematic variation of the TRGB discontinuity
occurs beyond the $10\arcmin$ radius and therefore the results for
fields P7--P9 yield the real distance estimate to M\,87.}\label{fig12}
\end{figure}

We present the results of all measurements in Table~\ref{fig4} and in Fig.~\ref{fig12}.
 Paired TRGB discontinuity values for each
radius are due to the use of two software packages---DAOPHOT and
DOLPHOT---for performing photometry in each field. The variation
of the TRGB discontinuity from $I = 27.1$ to~$26.9$ with the
distance from the center of the galaxy is immediately apparent in
Fig.~\ref{fig12}. At radii greater than $10\arcmin$ the
TRGB discontinuity value remains unchanged within the measurement
errors. We discuss the possible causes of such variation in section~\ref{sect:results}. But in any case the most
reliable distance estimates can be obtained from the results of
photometry in fields P7--P9.

We used the results of photometry in these fields to find the mean
distance modulus of M\,87 based on six measurements in three
fields and corrected for Galactic foreground extinction
\mbox{$(m-M) = 30.91\pm0.08$}, which corresponds to the distance
of $D = 15.39\pm0.57$~Mpc. Such a distance estimate puts the M\,87
galaxy to the nearest boundary of the cluster rather than into its
center, if only the cluster itself is not closer to us than \citet{Fouqu_etal:2001} believed.

\section{Distances to dwarf galaxies}
\begin{figure}[bpt!!!]
	\centering
\includegraphics[angle=0,
width=11cm,clip]{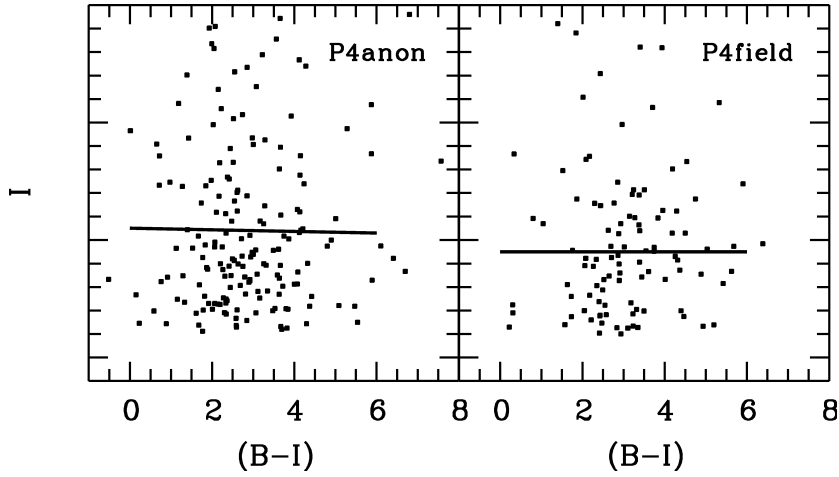} \caption{Color-magnitude
diagrams of the dwarf galaxy P4anon and the nearby background
area. The samples of the dwarf-galaxy and background-area stars
consist of 202 and 132 objects, respectively, i.e., red giants of
the dwarf galaxy make up one third of the samples. The position of
the TRGB discontinuity on the luminosity function of the  P4anon
galaxy indicates (Fig.~\ref{fig15}) that it is located in front of the
giant galaxy M\,87 and is its satellite.}\label{fig13}
\end{figure}

\begin{figure}[bpt!!!]
\centering
\includegraphics[width=0.75\columnwidth]{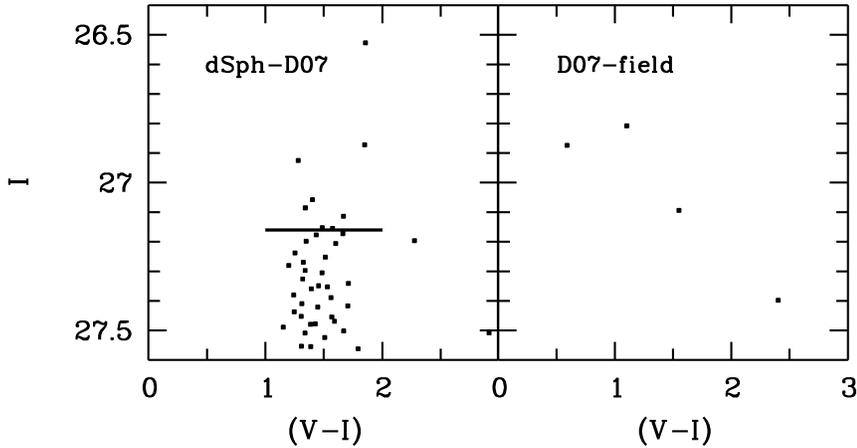}
\caption{Color-magnitude diagrams of the dwarf galaxy P9(dSph-D07)
and the nearby background area. The samples of the dwarf-galaxy
and background-area stars consist of 40 and four objects,
respectively. The value of the TRGB discontinuity at $I = 27.16$
indicates that this galaxy is 2~Mpc farther from us than  M\,87.}\label{fig14}
\end{figure}

Very faint galaxies projected onto M\,87 and resolvable into stars
can be seen in the images of fields P4 and P9 (Fig.~\ref{fig2}). When we
determine the distance to such a galaxy, the sample includes stars
of two galaxies: the dwarf studied and the huge  M\,87 background
system onto which  the dwarf is projected. To increase the
fraction of stars of the dwarf galaxy compared to that of stars of
the background
 M\,87 galaxy, we decreased the area of the sample for it to fit within the boundaries of the
dwarf galaxy. Furthermore, when determining the  TRGB
discontinuity we compared the luminosity function of stars of the
dwarf galaxy to that of the stars of the background area located
near  M\,87 and containing only its stars. The most challenging
case is that of the anonymous dwarf galaxy in field P4---P4
(anonymous) hereafter referred to as P4anon because of its small
size and small number of stars. The sample of stars of this galaxy
contains 202 objects, whereas the nearby background area contains
132 stars. Fig.~\ref{fig13} shows  the color-magnitude diagrams for these
areas. Despite the small number of stars, the TRGB discontinuity of
this galaxy can be determined quite reliably and its value differs
from the TRGB discontinuity of background stars, which can be
determined using a larger area rather than the small sample shown
in Fig.~\ref{fig1}. The dwarf galaxy in field P9 contains only 40 stars,
but the nearby comparison area contains only four stars (Fig.~\ref{fig14}),
and therefore the TRGB discontinuity is quite conspicuous and
measuring the distance to this galaxy  poses no problem.

\begin{figure*}[bpt!!!]
\centering
\includegraphics[width=0.8\textwidth]{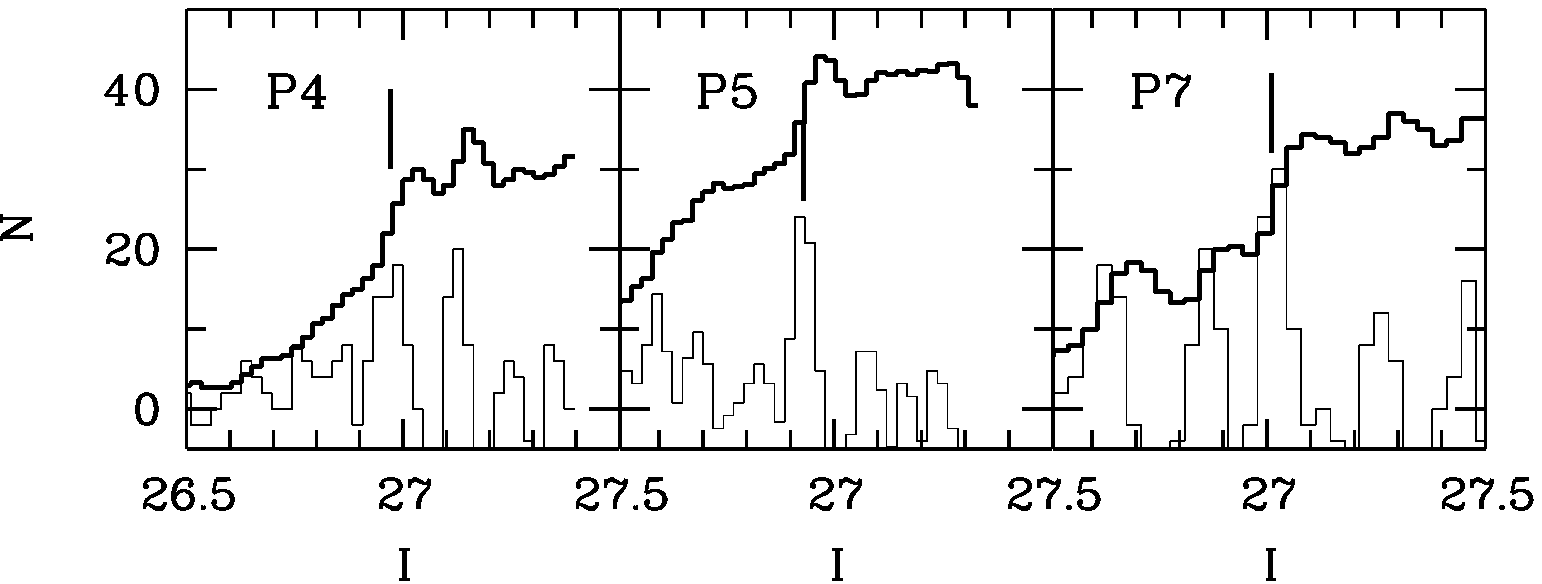}
\caption{Luminosity functions of the dwarf galaxies P4anon,
NGC\,4486A, and VCCA039. Designations are similar to those used in
Figs.~\ref{fig10}~and~\ref{fig11}. The values of the TRGB discontinuities differ
only slightly from that of  M\,87, i.e., these galaxies are
located at the same distance as M\,87 and are its satellites.}\label{fig15}
\end{figure*}

The galaxies  NGC\,4486a (field P5) and VCCA\,39 (field P7) have
much greater sizes than the dwarfs in fields P4 and P9, but their
central regions cannot be resolved into stars because of high star
density and high brightness. We plotted the star number density
distributions along the radius for these galaxies and determined
the radius $R_{\rm gal}$ at which the density of galaxy stars
becomes greater than the density of the background, which consists
of  M\,87 stars. After determining the $R_{\rm gal}$ radius we
selected stars for computing the luminosity function inside the
ring of radius $R < R_{\rm gal}$ and did not use the central
regions. Fig.~\ref{fig15} shows the resulting luminosity functions for
three galaxies whose distances have not been determined earlier via the TRGB
method. Table~\ref{tab4} lists the TRGB discontinuities ($I_{\rm TRGB}$)
for the galaxies P4anon, NGC\,4486a, and VCCA\,39. The measured
distances to these galaxies are equal to 15.6, 15.6, and 16.1~Mpc,
respectively. The distance measurement accuracy directly depends
on the accuracy of the determination of the location of the TRGB
discontinuity and that of the equations used to transform this
position into linear distance units. The accuracy of
transformation equations~\citep{Lee_etal:1993} and that
of the determination of the TRGB
discontinuity are equal to $I = 0\fm10$ and of about $0\fm03$,
respectively. The transformation equations from instrumental
magnitudes to the Cousins system are accurate to $0\fm02$ and the
accuracy of the computation of the PSF profiles of stars is about
the same. The combined contribution of all measurement errors is
$0\fm16$, which is equivalent to the 1.3~Mpc accuracy of the
distance estimates for each galaxy.

The distances to  P4anon, NGC\,4486a, and VC\-CA\,39 differ little
from the distance to M\,87, i.e., they are satellites of the giant
M\,87 galaxy. The fourth dwarf galaxy that located in field P9
(dSph-D07) is known since long and earlier distance estimates are
available for
it:\,$D\!=\!16$\,Mpc\,\citep{Williams_etal:2007a},\,\mbox{$D\!=\!17.6\!\pm\!1.4$\,Mpc}\,\citep{Durrell_etal:2007},
\mbox{$D\!=\!18.3\!\pm\!1.4$}\,Mpc\,\citep{Jang_Lee:2014}.
Our distance measurement yielded \mbox{$D = 17.1\pm1.5$~Mpc} with
the red-giant metallicity of \mbox{$[Fe/H] = -1.8$}, which agrees
with earlier measurements within the errors. This dwarf spheroidal
galaxy is almost 2~Mpc farther than M\,87 and belongs to the group
of the more distant galaxy M\,86 with \mbox{$D = 18.5\pm0.5$~Mpc}
~\citep{Tikhonov:2017}.

\begin{table*}
	\begin{center}
\caption[]{TRGB discontinuity values in fields P1--P9 around M\,87
and in dwarf galaxies}\label{tab4}
\medskip
 
\begin{tabular}{l|c|c|c|c|c|c|c|c|c} \hline\noalign{\smallskip}
 ~~~~~Object & P1  & P2  & P3 & P4 & P5 & P6 & P7 & P8 & P9  \\
\hline\noalign{\smallskip}
M\,87 (DAO)*& --  &27.12&27.03& 27.07& 27.01& 26.88& 26.99& 26.85& 26.89 \\
M\,87 (DOL)*&27.10& --  & --  & 27.02& 27.05& 26.97& 26.86& 26.95& 26.92 \\
P4anon   & --  & --  & --  & 26.97&  --  &  --  &  --  &  --  & --    \\
NGC\,4486a   & --  & --  & --  &  --  & 26.93&  --  &  --  &  --  &  --   \\
VCCA\,039  & --  & --  & --  &  --  &  --  &  --  & 27.01&  --  &  --   \\
dSph-D07 & --  & --  & --  &  --  &  --  &  --  &  --  &  --  & 27.16 \\
\hline\noalign{\smallskip}
\end{tabular}
\begin{tabular}{lp{150mm}}
* based on the results of  DAOPHOT (DAO) and DOLPHOT (DOL) photometry
\end{tabular}
 \end{center}
\end{table*}

\section{Results and discussion}
\label{sect:results}
Our aim was to investigate the parameters of stars in the area
between the M\,86 and M\,87 galaxies and make conclusions about
their origin. We concluded that the study has to be extended to
investigate the structure of the  M\,87 subsystem and determine
the space position of this galaxy in Virgo cluster.

While addressing these tasks we demonstrated the efficiency of the
star count method. This is practically the only method that allows
studying the morphology and stellar composition of the faint and
extended periphery of a galaxy. We used the technique of star
counts for several fields located at different distances from the
center of M\,87 to demonstrate the variation of the  number
density of AGB stars and red giants of various luminosities and
metallicities along the galactic radius (Figs.~\ref{fig6} and~\ref{fig7}). We
established that the most luminous AGB stars show the largest
gradient of density decrease, and that gradient decreases `when
passing  to less luminous stars. A dependence was found between
the metallicity of red giants and the gradient of the variation of
the star number density along the galactic radius. The results
obtained explain the stellar composition of the field between the
galaxies M\,86 and M\,87 and suggest that stars of this field (P9
according to our numeration) predominantly belong to the periphery
of  the M\,87 galaxy. So, there are no grounds to consider them to
be likely intergalactic stars, and even if they are part of the
sample, they still do not affect its parameters because of their
small number. The fact that stars of M\,87 are found at a distance
of 190~kpc from its center indicates that the  size of the stellar
halo is not smaller than  this value. It is evident from the
images obtained by~\citet{Mihos_etal:2005,Mihos_etal:2017}
that the halo of M\,87 has a shape of a regular ellipse, and
therefore the detection of halo stars at a single location in this
ellipse determines the size of the entire halo up to this radius.
In the diagram in Fig.~\ref{fig5} many stars  that belong to the
outermost field of M\,87 can be seen. Given the smooth decrease of the stellar
number density, we can conclude that the halo extends beyond this
field located at a distance of 190~kpc from the center of the
galaxy. Hence the size of the halo  is greater than 190~kpc.

The conclusions reached in this paper and the results of our
earlier studies~\citep{Tikhonov:2017} show that stars in
all the Virgo cluster fields that we studied belong to the
peripheries of neighboring galaxies and are not intergalactic
stars as a number of studies have
suggested~\mbox{\citep{Caldwell:2006,Williams_etal:2007a}.}
Apparently, the process of stripping of
galaxies~\citep{Rudick_2006} is not an efficient
mechanism for generating intergalactic stars in clusters.

Determining the distance to the  M\,87 galaxy via the TRGB method
revealed the variation of the TRGB discontinuity with increasing
distance from the center of the galaxy (Fig.~\ref{fig12}). This effect can
be explained by the variation of the morphological composition of
the star sample: fields closer to the center of the galaxy contain
red giants with higher metallicity, and this is what cases the
systematic shift of the TRGB discontinuity as observed in the
diagram in Fig.~\ref{fig12}.

There is another possibility, less likely cause of the observed
dependence in Fig.~\ref{fig12}. During its entire lifetime, the giant  M\,87
galaxy has cannibalized quite a few dwarf galaxies. Dust of these
galaxies may have formed a weak halo around M\,87 causing light
absorption and shifting the TRGB discontinuity. This would result
in fictitious increase of the  distance as we go to samples
located closer to the center of the galaxy. This effect can
probably also explain the systematically greater distance to M\,87
determined using the method based on nova stars, which explode
more often in central regions of the galaxy. Extensive studies of
novas in
M\,87~\citep{Shara_etal:2016,Shara_etal:2017}
will possibly resolve the problem of measuring the distance via
this method. We believe that the most accurate  M\,87 estimates
are those provided by measurements performed via the TRGB method
applied to fields~\mbox{P7--P9} at the periphery of the galaxy.

Our distance estimate to M\,87 puts this galaxy to the forefront
of the cluster if only the entire cluster is not located closer
than~\citet{Fouqu_etal:2001} believed. The
asymmetric position of  M\,87 in the cluster is not unexpected.
For example, all measurements based on planetary nebulas yielded
small distances to M\,87  (NED). Recent measurements of this type
include: \mbox{$D =
14.1$~Mpc}~\citep{Longobardi_etal:2013}, $D =
14.5$~Mpc~\citep{Longobardi_etal:2015}.

Three dwarf galaxies located near  M\,87 that happen to be in the
fields studied yield approximately the same distances as  M\,87,
thereby indirectly corroborating the correctness of our distance
measurements to M\,87.

M\,87 does not need to be at the center of the cluster. It is by
no means the brightest galaxy in the cluster. The elliptical
galaxy M\,49, which is located at the cluster boundary, is
brighter than M\,87. It has been repeatedly shown that Virgo
cluster is a loose assembly of individual groups of
galaxies~\mbox{\citep{Neilsen_Tsvetanov:2000,
Fouqu_etal:2001, Solanes_etal:2002,
Mei_etal:2007}}, which near future may form a
regular cluster of galaxies.

The question about the location of individual groups of galaxies
inside Virgo cluster remains open because of the lack of accurate
distance measurements of other massive cluster galaxies besides
M\,87. Numerous but not accurate measurements made using the
Tully--Fisher (TF) or surface brightness fluctuations (SBF)
methods cannot replace few but accurate measurements made using
the TRGB method. Most of the members in the cluster are lenticular
and elliptical galaxies where a large fraction of visible stars is
represented by red giants and therefore the TRGB method proves to
be practically the only tool for accurate distance measurement.
However, for this we need Hubble Space Telescope images taken
with $F814W$-band exposures no shorter than 5000~s in fields
located at the periphery of the galaxies rather than at its
center.

\normalem
\begin{acknowledgements}
	
Based on observations with the NASA/ESA Hubble Space Telescope,
obtained at the Space Telescope Science Institute, which is
operated by AURA, Inc. under contract No. NAS5-26555. These
observations are associated with proposals 10131, 10543, 12532,
12989 and 13731.

\end{acknowledgements}

\end{document}